\documentclass[12pt]{article}
\usepackage[english]{babel}
\usepackage[dvips]{color}
\usepackage{amsmath,epsfig,calc}

\definecolor{red}{cmyk}{0.1,1,0.8,0}
\definecolor{green}{cmyk}{1,0,1,0.4}

\begin{document}
\selectlanguage{english}

\begin{titlepage}

  \vspace{3cm}
  \begin{center}

    \stepcounter{footnote}
    \renewcommand{\thefootnote}{\fnsymbol{footnote}}
   {\Large \textbf{Causality Aspects of the Parton Cascade Approach to
                   Ultarelativistic Heavy Ion Reactions}
    \footnote{Poster presented at \emph{Quark Matter 2004}, January 13,
              2004
             }
   }

   \vspace{0.5cm}
   {\large C.C.\ Noack \\ }
   \vspace{0.3cm}
   \emph{Institut f{\"u}r Theoretische Physik
     Universit{\"a}t Bremen,\\ D--28334 Bremen, Germany\\
     e-mail: \texttt{noack@physik.uni-bremen.de\\ }}
 \end{center}
\end{titlepage}

    \setcounter{footnote}{0}
    \renewcommand{\thefootnote}{\arabic{footnote}}
\section*{\textcolor{blue}{Introduction} }

Parton cascade codes that take a space-time approach to model the
microscopic processes
\footnote{Prominent examples are \textbf{VNI} and \textbf{pcpc}; cf.\
          the OSCAR archive at \\
          \texttt{http://nt3.phys.columbia.edu/people/molnard/OSCAR/} .
         }
of an ultrarelativistic heavy-ion interaction are -- in spite of their
QCD bells and whistles -- by necessity based on some kind of
\emph{classical many-particle dynamics}.

\subsection*{\textcolor{red}{Problems:} }
\begin{itemize}\item space-time-based cascade models suffer from the consequences of
      the
      \begin{center} \textcolor{green}{No-Interaction Theorem (NIT)}
                                           \cite{Cur63}: \\[1ex]
          \fbox{\textcolor{green}{\parbox{\textwidth*2/3}{\centering
                The only consistent many-particle \\
                Hamiltonian theory that is Poincar{\'e}-covariant \\
                is that of a system of free particles.} } }
      \end{center}
   \item the procedure to determine the sequence of the binary parton
      interactions (``SBPI'') -- an essential aspect of space-time-based
      cascade models -- is by necessity an artificial and ad-hoc feature
      of these codes.
  \end{itemize}

The only way to circumvent the NIT is to loosen its assumptions:
\begin{enumerate}\item forget about Poincar{\'e} covariance \hfill
      $\Longrightarrow$\qquad \textbf{VNI}  \\[1ex]
      In this approach the \textsc{SBPI} depends on the initially chosen
      frame of reference. Einstein causality remains a problem.
   \item introduce a many-times formalism, e.g.\ by formulating the model
      in $8N$-dimensional phase space ($N$ is the particle number)
      \footnote{We have shown previously that for $N=2$ all
                known Poincar{\'e}-covariant formalisms are equivalent to
                ours (cf.\ \cite{Pet94}).
               } \\
      \hspace*{\fill} $\Longrightarrow$\qquad \textbf{pcpc} \\[1ex]
      The Poincar{\'e} covariance of this model seems to guarantee Einstein
      causality; the SBPI of the code, however, deserves closer
      scrutiny.
\end{enumerate}

\section*{\textcolor{blue}{Basic Covariant Structure of \textbf{pcpc}} }

\textbf{pcpc} is a hybrid of a \emph{classical} dynamics approach that
governs the evolution of the system between binary parton interactions,
and a parton interaction model with QCD ingredients (parton distribution
functions, pQCD cross sections and `DGLAP evolution')
\footnote{For details, cf.\ \cite{Pet94,Boe00}}.
\begin{itemize}\item The dynamical evolution of the system is parametrized by a
      Poincar{\'e} scalar, $s$ (in contrast to the usual $t$, the proper
      time of an external observer, as measured in some external frame)
   \item The phase-space variables of the $N$ partons (quarks, gluons) are
      covariant 4-vectors ${x_i\,}^\mu(s)$, ${p_i\,}^\mu(s)$ . Between
      binary interactions $N$ is fixed, but can (and does) vary due to
      parton creation in the QCD-governed parton interactions

   \item The ($8N$-dimensional) interaction term of the
      Poincar{\'e}-\emph{invariant} \linebreak
      `Hamiltonian' depends on the following
      Poincar{\'e}-invariant `4-distances'
      \[ {d_{ij}\,}^2(s) := -{\hat{x}\,}^2 = -(x\hat{x}) = -(\hat{x}x)\]
      \[ \hat{x}\,^\mu := x\,^\mu - \frac{(xp)}{p^2}p\,^\mu \]
         [\, $x\,^\mu(s):={x_i}\,^\mu(s) -{x_j}\,^\mu(s), p\,^\mu(s):=
         {p_i}\,^\mu(s) +{p_j}\,^\mu(s)$ are the relative 4-distance and
         total 4-momentum of particles $i$ and $j$ \,]

      The physical reasons why in Poincar{\'e}-covariant dynamics the
      interactions can only depend on these ``orthogonal projections''
      ${\hat{x}}^\mu$ have been given before \cite{Pet94}.
   \item Binary parton interactions occur at the $s$ determined by
      ${d_{ij}}^2(s)$ being at a minimum
      \footnote{The interaction term in the Hamiltonian thus can be
                thought of as a sum of $\delta$-functions.
               }
      (in the cms of partons i, j, ${d_{ij}}^2$ is the minimal
      3-distance of approach)
   \item Between interactions, partons move along free trajectories:
      \[ {x_i\,}^\mu(s) = \frac{{p_i\,}^\mu(s)}{m_i}(s -s_0)
             +{x_i\,}^\mu(s_0) \]
\end{itemize}

\section*{\textcolor{blue}{Logical flow in \textbf{pcpc}} }

In contrast to most other cascade codes, the `time step' in
\textbf{pcpc} is not a computational artifice, but is given by the
formalism itself:
\begin{itemize}
\item if there were only 2 partons, their interaction would occur at the
   value of the evolution parameter $s$ at which they would reach their
   minimal ${d_{ij}\,}^2(s)$ if they were indeed alone in the world,
\item so a `time table' is kept, containing, for every pair $i,j$ of
   partons, its potential minimal approach ${d_{ij}\,}^2(s)$ (i.e.\
   their minimal approach if $i,j$ were the only partons in the whole
   system), and the corresponding $s$,
\item then this table is searched for the \emph{smallest} (`earliest') $s$.
   At this $s$ the potential interaction of the corresponding pair
   \emph{will actually occur} (because all other potential interactions
   are `later'),
\item the interaction of the pair $i,j$ will change the world lines of
   these two partons (and possibly create further partons). Therefore,
   the time table is updated, with new values of ${d_{ij}\,}^2(s)$ for
   all pairs involving either parton $i$ or parton $j$ (or the newly
   created partons); and the code loops.
\end{itemize}

\bigskip
Thus the code follows exactly the sequence of binary parton interactions
as parametrized by the monotonically increasing
\emph{Poincar{\'e}-invariant} evolution parameter $s$. The logic is
summarized in the following flow diagram:
\begin{center} \includegraphics[scale=.75]{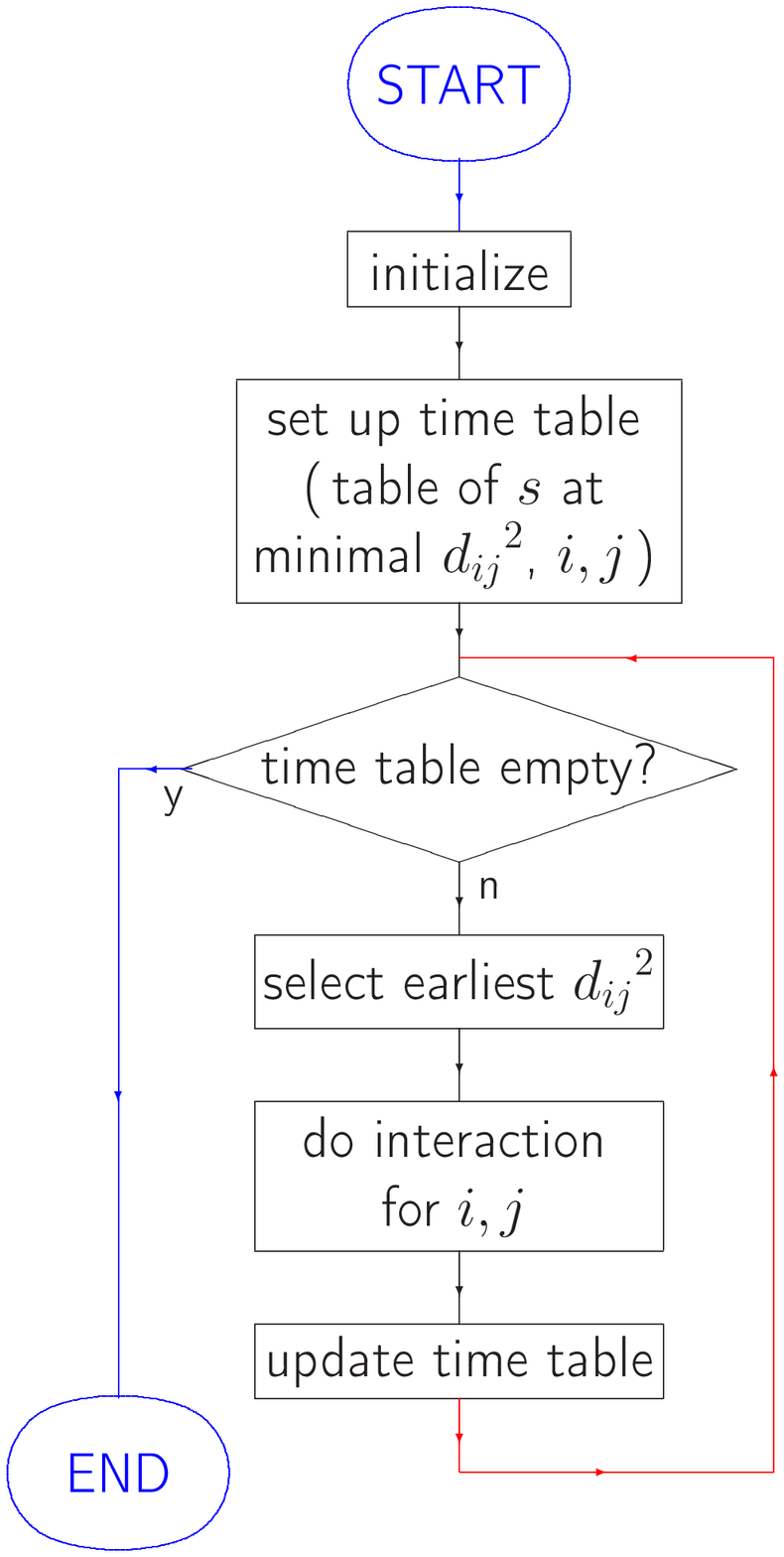} \end{center}

\section*{\textcolor{blue}{Do the parton interactions preserve Einstein
          causality?} }

Prima facie this does not seem to be so: the 4-vector ${x_i\,}^\mu
-{x_j\,}^\mu$ is space-like, and so no signal can be transmitted between
events $x_i$ and $x_j$. But this argument is fallacious: while it would
be correct in the frame work of a $6N$-dimensional phase space formalism
and physical observer time, it does not follow in a many-times formalism
with an $8N$-dimensional phase space.

\medskip
Furthermore, it must be remembered that the representation of the (QCD)
physics of the heavy-ion reaction in terms of cross sections, parton
distribution functions etc.\ is tantamount to the description of
intrinsically quantum processes in a \emph{classical} terminology. But
as long as we refrain from trying to look \emph{inside} an individual
binary parton interaction with classical concepts, Einstein causality is
not infringed upon by such a model (for details on this point cf.\
\cite{Pet94}\,).

\section*{\textcolor{blue}{Is Einstein causality preserved between
          binary interactions?} }

In terms of the cascade picture, a heavy-ion reaction can be seen as one
or several disjoint graphs of connected particle world lines. In every
connected subgraph, the Poincar{\'e}-covariant dynamics guarantees Einstein
causality.

\medskip
Separate disjoint subgraphs, however, can have no causal connection. But
by construction, no signals are transmitted between them. Imagining a
full quantum-mechanically description of such a system, unconnected
subgraphs would correspond to subamplitudes which would simply be
multiplied in obtaining the total amplitude.

\section*{\textcolor{blue}{Are the initial parton \emph{positions}
          critically important?} }

There is no physically convincing argument for how to set the time
components of the initial parton 4-vectors $a_i :=
{x_i\,}^{\mu=0}(s=s_0)$; so these must be fixed phenomenologically with
some arbitrary prescription. In \textbf{pcpc} they are all set to zero.

Does this imply the choice of a particular initial frame of reference,
thus invalidating covariance, or spoiling the Einstein causality of the
model? The answer is
\begin{center} \textcolor{green}{\textbf{\large NO!}} \end{center}
To see this, suppose that we \emph{do not} set the initial $a_i=0$, but
retain them as free (arbitrary) parameters. We would then find the
minimum of the ${d_{ij}\,}^2$ to be formally dependent on the $a_i$. But
since for any two time-like 4-vectors the invariant quantity
$(x\hat{y})$ is simply $\left.\rule[-1.5ex]{0pt}{4ex}
-\vec{x}\cdot\vec{y}\,\right|_{\text{cms}}$, we find that the
${d_{ij}\,}^2$ are actually \emph{independent of the} (time components
of the) \emph{initial parton positions}. It follows that the parton
interactions are implemented in a Poincar{\'e}-covariant way, even though
their sequence (SBPI) \emph{does} depend on how the initial parameters
are chosen.

\section*{\textcolor{blue}{Conclusions} }
To sum up, causality is \emph{not an issue} in discussing the physical
validity of a cascade model that uses a space-time approach to the
dynamical evolution (provided the model is Poincar{\'e}-covariant).

\bigskip
It is, however, important to realize that in such codes the sequence of
binary parton interactions (SBPI) is to be considered an essential
\emph{part of the model}, and that it is \emph{necessarily
phenomenological} in character. In Poincar{\'e}-covariant cascade codes,
such as \textbf{pcpc}, the SBPI (although remaining a phenomenological
prescription) is independent of choice of the frame of reference in
which the code is run.

\medskip
The difficulties with Einstein causality incurred by non-covariance of
the SBPI have been discussed many years ago \cite{Kod84}. In contrast to
the situations described in that paper, Einstein causality is preserved
in a Poincar{\'e}-covariant model such as \textbf{pcpc}, both for the
individual binary interactions and for the dynamic evolution of the
system as a whole.


\begin{thebibliography}{Cur63}
\bibitem[CJS63]{Cur63}
  D.\ Currie, T.\ Jordan, and E.\ Sudarshan,
  Rev.Mod.Phys.\ \textbf{35}, 350 (1963)

\bibitem[KDCDN84]{Kod84}
  T.\ Kodama, S.\ Duarte, K.\ Chung, R.\ Donangelo and R.\ Nazareth,
  Phys.Rev.\ \textbf{C 29}, 2146 (1984)

\bibitem[PNB94]{Pet94}
  G.\ Peter, C.\ Noack, and D.\ Behrens,
  Phys.Rev.\ \textbf{C 49}, 3253 (1994).

\bibitem[BMGMN00]{Boe00}
  V.\ B{\"o}rchers, J.\ Meyer, S.\ Gieseke, G.\ Martens and C.C.\ Noack,
  Phys.Rev.\ \textbf{C 62}, 064903 (2000)

\end{thebibliography}
\end{document}